# Experimental demonstration of wakefield effects in a THz planar diamond accelerating structure

S. Antipov[1,2], C. Jing[1,2], A. Kanareykin[1], J.E. Butler[1], V. Yakimenko[3], M. Fedurin[3], K. Kusche[3], W. Gai[2]

[1]Euclid Techlabs LLC
[2]Argonne Wakefield Accelerator Facility, Argonne National Laboratory
[3]Accelerator Test Facility, Brookhaven National Laboratory

## Abstract

We have directly measured THz wakefields induced by a subpicosecond, intense relativistic electron bunch in a diamond loaded accelerating structure via the wakefield acceleration method. We present here the beam test results from the diamond based structure. Diamond has been chosen for its high breakdown threshold and unique thermoconductive properties. Fields produced by a leading (drive) beam were used to accelerate a trailing (witness) electron bunch which followed the drive bunch at a variable distance. The energy gain of a witness bunch as a function of its separation from the drive bunch describes the time structure of the generated wakefield.

## Paper

High gradient dielectric – loaded accelerating structures (DLA) have received significant attention in recent years [1-4]. Such devices are rather simple to fabricate. Typical examples would be a dielectric tube inserted into a conducting jacket or a rectangular waveguide loaded with dielectric plates. DLAs are compact; magnetic correctors in the form of a solenoid or series of quadrupoles can be placed around the structure. More importantly, dielectrics exhibit higher breakdown thresholds [2] and they are easily scalable to THz [2 - 4] and even higher frequencies where they have lower total structure losses compared to metals. Dielectric structures are also less sensitive to the single bunch beam break-up (BBU) instabilities [5].

A dielectric accelerating structure can be externally powered. This requires the design and fabrication of a coupler. Alternatively a DLA can be used as a wakefield structure [1], in which the accelerating field is generated by a low energy high charge, drive bunch. Such a scheme is simple and attractive. The experimental results reported in this paper were obtained via the wakefield technique. The experiment was conducted at the Accelerator Test Facility (ATF) of the Brookhaven National Laboratory. It is also worth pointing out that these dielectric structures can also be used as intense sources of THz radiation [3, 4]. CVD (chemical vapor deposition) diamond has been proposed [6, 7] as a dielectric material for wakefield DLA structures [1-3]. It has a very low microwave loss tangent, the highest available thermal diffusivity and a high RF breakdown field. The large secondary emission from the as manufactured CVD diamond surface can be dramatically suppressed by diamond surface dehydrogenation or oxidation [8]. CVD diamond has already been routinely used on an industrial basis for large-diameter output windows of high power gyrotrons, and is being produced industrially in reasonable quantities.

In this paper we present results from experimental studies of a rectangular waveguide loaded with polycrystalline CVD diamond plates as an accelerating structure. In this experiment a drive (high charge) electron beam was launched through the structure creating a wakefield (Cherenkov radiation) following behind it. A smaller charge, (witness) beam was launched with variable delay with respect to the drive beam and experienced acceleration or deceleration depending on the delay. This measurement effectively maps the wakefield produced by the drive beam. The work presented here is the mapping of a wakefield in a dielectric device in the THz frequency range at the scale of about one wavelength. It should be noted especially that in this experiment diamond was used as a loading material for the wakefield dielectric based accelerating structure.

In the experiment we used a 75 micron thick polycrystalline diamond plates loaded in a 6 cm long waveguide (figure 1). The beam gap is 200 microns (figure 1b). This structure yields a wakefield dominated by a $TM_{11}$ – like mode with 1200 micron wavelength (0.25 THz). The ATF drive beam is very short and also excites higher order modes, hence the wake is not a pure sine wave (see figure 2b).

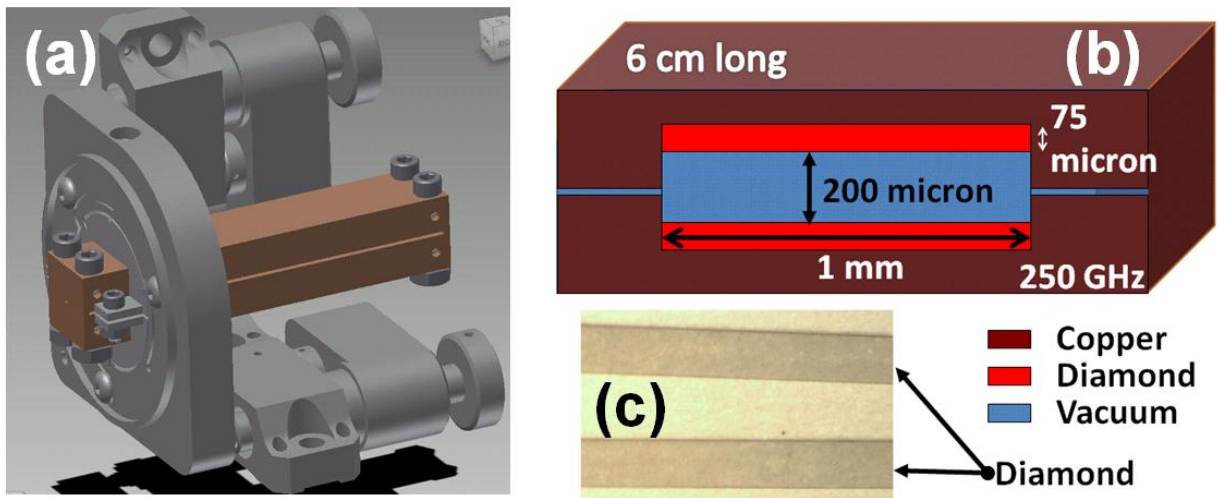

Figure 1. Experimental geometry. (a) structure mounted in the motorized holder. (b) waveguide schematic with dimensions. (c) photo of the diamond strips taken out of the structure *after* the experiment. Note that no damage to the diamond is observed.

The subpicosecond drive and witness beams are produced by the technique described in [9]. A beam with a correlated energy spread (for example with the head of the beam having lower energy than the tail) passes through a dogleg – a dispersionless translating section that consists of two bending dipoles whose bend angles are equal in magnitude but opposite in sign. Focusing optics are located between the dipoles. A mask is placed between the dogleg dipoles where the beam transverse size is dominated by the correlated energy spread. After the second dipole magnet the transverse pattern introduced by the mask becomes the longitudinal charge density distribution. In our case the mask was motorized and allowed creation of a drive beam followed by a witness beam at a variable delay. There is a one to one correspondence between the mask pattern and the beam's longitudinal distribution after the second dipole magnet. Therefore with proper calibration the image of the beam directly after the mask represents longitudinal beam distribution (for both drive and witness bunch – figure 2a, b). This calibration is done via coherent transition radiation (CTR) interferometry. In this technique two test beamlets are created by the mask and recorded on a phosphor screen after the mask. In particular the distance between the beamlets is measured on the phosphor screen. Then after the second dipole the distance (along z) between these two beamlets is measured by interferometric manipulation of transition radiation signal which the beamlets emit passing through a thin CTR foil. Wavelengths longer than the beams are emitted coherently and carry information about the bunch lengths [10, 11]. Transition radiation is sent to an interferometer and the signal is recorded by a helium-cooled bolometer [4, 9]. From this measurement the distance between beamlets was determined and the phosphor screen image was calibrated. In our experiment we measured drive beam longitudinal size to be 320 microns, and the witness beam was about 100 microns long. Figure 2a shows the image on the phosphor screen after the mask and figure 2b shows the longitudinal current distribution obtained by this measurement and used in the theoretical wakefield computation. In this case we neglected the dispersion of the beam while it was being delivered to the experimental area where the diamond structure was installed. The dispersion effect was estimated by simulation (program MAD (Methodical Accelerator Design) [12]) to be less than 5%.

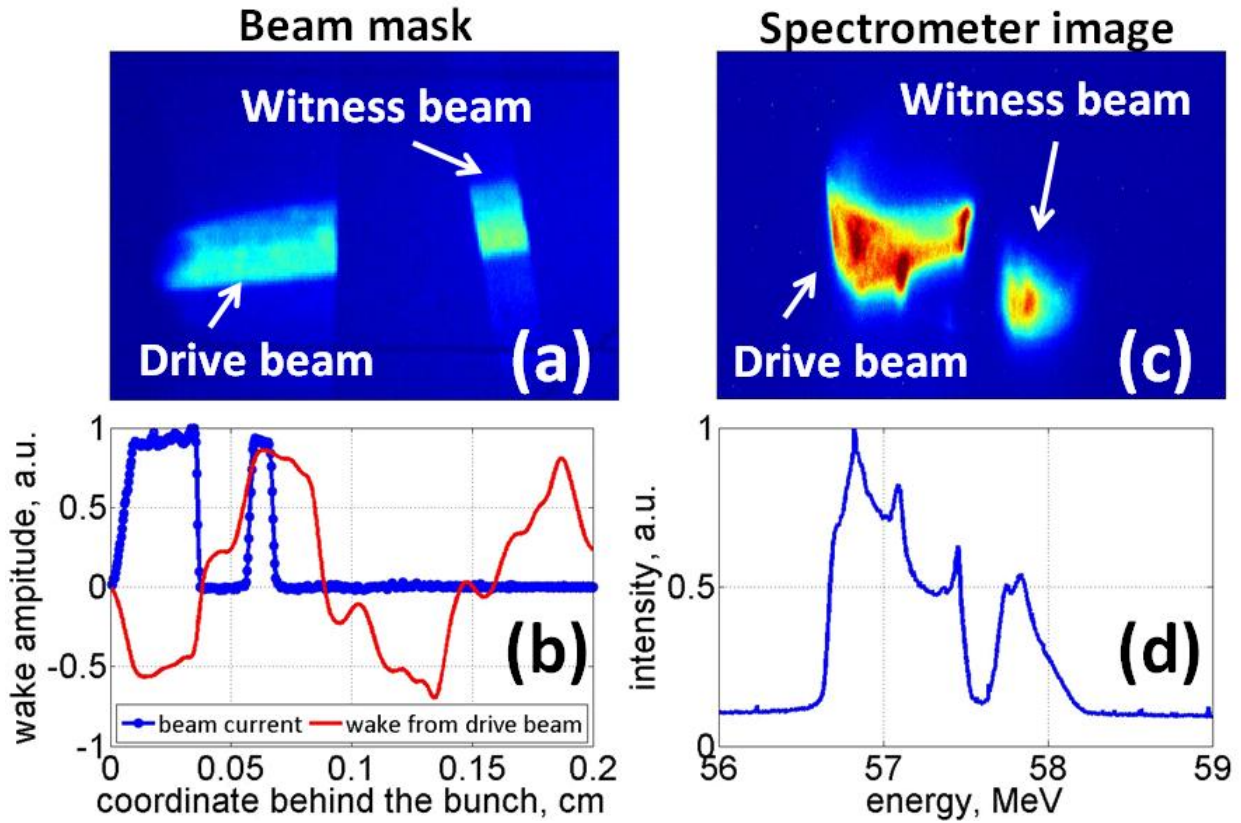

Figure 2. (a) Image after the beam mask. (b) Drive and witness beam longitudinal current profile obtained from the image (a) and wakefield produced by the drive beam only. (c) Spectrometer image of the witness beam 285 microns behind the drive beam. (d) Spectrometer projection. Energy spread of the drive beam is 0.855 MeV.

It is worth noting that the shaped beam still carries a correlated energy spread, in our case the head (drive) beam having higher energy than the witness beam (tail). This has two consequences: First, this means that in our case of no dispersive beam optics downstream, the spectrometer image will provide longitudinal distribution information: we will be able to see the drive beam separate from the witness beam (figure 2c, d). Information obtained by delivering the beam to the spectrometer without passing it through the DLA structure can be used as a double check for longitudinal beam distribution measurement and calibration. Second, when a new pair of drive and witness beams is produced by adjusting the motorized mask, the new witness beam has higher energy then the previous witness beam due to the correlated energy spread required for this beam shaping technique. It is essential to calibrate this effect out, because the wakefield mapping procedure relies on the measurement of witness beam energy gain or loss as a function of the drive – witness beam delay.

For theoretical analysis of the wakefield measurements we took the longitudinal profile of the beam obtained via CTR interferometry calibration of the beam image downstream of the beam shaping mask described above. The transverse beam distribution was obtained from the image on the phosphor screen directly in front of the structure. We fitted the recorded transverse profile by a gaussian distribution with $\sigma_x$ = 50 µm, $\sigma_y$ = 350 µm. We numerically calculated the longitudinal wakefield produced by a beam with this current distribution. This is done indirectly by summing over waveguide parameter r/Q (normalized shunt impedance calculated for the particle passing through a certain position ($x_0$, $y_0$) in the waveguide cross section) for each of the structure's modes [13].

$$E_{0i} = \frac{q\omega_i}{2}\left(\frac{r_{x_0,y_0}}{Q}\right) = \frac{q}{2}\left(\frac{|E_z|_{x_0,y_0}}{U}\right) \tag{1}$$

(The subscript *i* indicates mode number, ω – mode frequency, *r/Q(x₀,y₀)* is calculated for the particle passing through a certain position ($x_0$, $y_0$) in a waveguide cross section, Q – quality factor for the mode, *q* – particle charge, *U* – stored energy.) A particle with coordinates ($x_0$, $y_0$, $z_0$) excites the i-th mode at the following amplitude:

$$E_i = E_{0i}(x_0, y_0) \cdot \sin\left(\frac{\pi x_0}{w} + \frac{\pi}{2}\right) \cdot \cos(k_{yi}^0 y) \cdot \cos\beta_i(z - z_0) \tag{2}$$

Here *w* is the width of the rectangular waveguide, $k^0_{yi}$ is a transverse propagation constant, and *β* is the longitudinal propagation constant.

Then a summation of modes (2) is performed across the waveguide modes and across particles in the beam considering both transverse and longitudinal distributions. Summation across modes can be limited to the first ten or so modes as the contribution from the higher frequency mode becomes insignificant. The resulting wake is dominated by the $TM_{11}$ – like mode at 0.25 THz with some higher frequency components present (figure 2b).

Due to the relatively large longitudinal size of the witness beam its head and tail sample different fields. In most cases this results in an increase in the witness beam energy spread. The witness beam can split in two as well (figure 2c, d). We traced the centroid of the witness beam when post-processing the results (figure 3). The drive beam also creates a wakefield inside itself. This field changes the energy spectrum of the drive beam. We observe this effect as a distortion of the drive beam shape from its original rectangular form (figure 2c, d). This effect can be used for energy modulation of beams with various shapes and is the subject of a separate experiment and publication [14].

The spectrometer image (figure 2c, d) is calibrated by varying the current in the spectrometer magnet. The relative change in witness beam energy depending on the drive – witness separation is measured on the spectrometer. Compared to the beam energy, when it does not go through the structure, we observe energy gain and loss in the interval from -0.6 to +0.65 MeV for a witness beam following a drive beam through the DLA structure. This yields a 10.8 MV/m gradient for a 6 cm structure. The measurement accuracy was limited by the spectrometer resolution ($\sigma$ = 0.028 MeV), estimated using the sharpest features obtained on the spectrometer by measuring the FWHM (full width half maximum) and dividing it by 2.355 to obtain the equivalent Gaussian width. Beam stability was also crucial for the relative energy measurement. The observed jitter of the spectrometer image corresponded to 0.015 MeV. Finally, the tracking centroid of the witness beam could be determined with accuracy equal to the standard deviation of its energy spread. Figure 3 shows the measured energy change (diamonds) as a function of drive – witness separation with error bars.

In the experiment we were able to achieve only a 70% transmission rate for about a 70 pC drive beam. Due to loss of charge it is hard to compare the theoretical model with the measurement magnitude-wise. We scale the charge in simulation to get theoretical wakefield to correspond to observed energy gain. From this scaled wakefield we obtain theoretical energy gain by integrating the accelerating field over the longitudinal charge distribution of the witness beam and computing the centroid of the resulting energy distribution. The shape of the theoretical energy gain / loss agrees well with the shape obtained in the experiment. The maximum separation between the drive and witness beams is limited by the maximum possible size the beam can be dispersed transversely by running off-crest in the accelerating section. We were able to sweep about 1 mm axial distance between drive and witness beams.

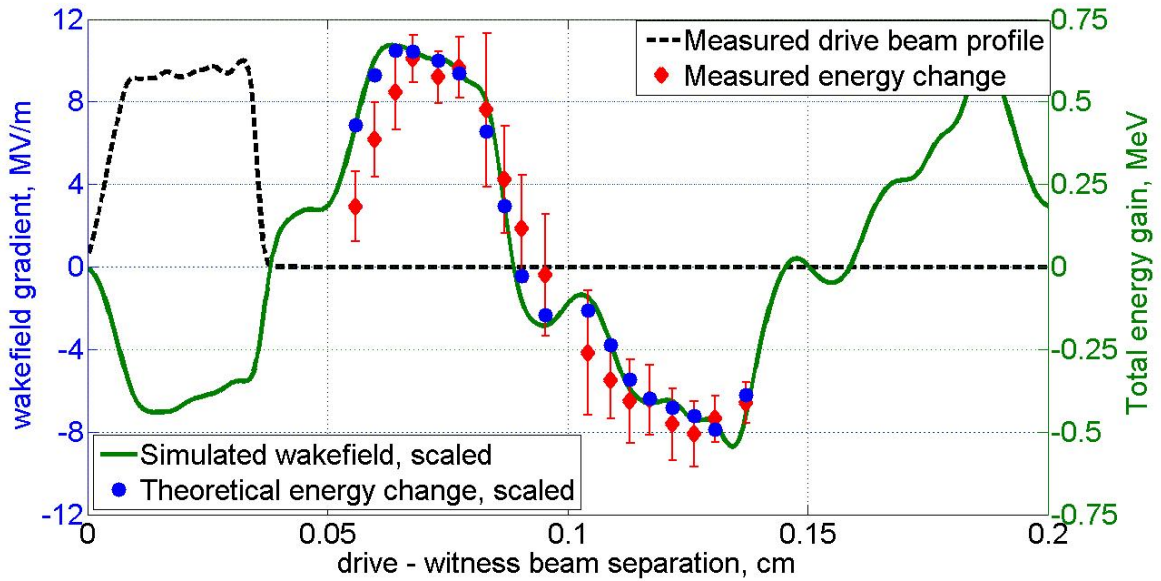

Figure 3. Energy gain of the witness beam as a function of separation from the drive beam. The shape of the experimentally mapped wake (diamonds) agrees well with the theoretical (scaled) prediction (circles) based on the wakefield (solid line) calculated for the drive beam current distribution (dashed line).

During the beam alignment the diamond plates were directly exposed to the electron beam. Besides the wakefield measurement of the diamond based DLA structure we decided to test the effect of the 57 MeV electron beam on a diamond polished surface. We performed inspection of the material after the experiment. Microscopic examination of the polycrystalline diamond plates used in the ATF experiment showed no visible damage to the diamond material from the 57 MeV electron beam exposure. Figure 1c shows a photo of two diamond plates removed from one side of the structure after cleaning off the residual epoxy (Varian torr seal) used to hold the diamond plates in the structure. There is no evidence of graphitization, amorphization, or etching of the diamond either along the bulk of the material or at the joints (edges) of the plates where high field gradients would be expected.

In conclusion we have directly measured wakefield acceleration / deceleration in a diamond loaded dielectric accelerating structure. The witness beam was accelerated in the field produced by a subpicosecond drive beam. By sweeping the separation between the drive beam and the witness beam and energy measurements of the witness beam a 0.25 THz frequency wakefield was directly sampled at the scale of about one wavelength.

# References

[1] W. Gai, P. Schoessow, B. Cole, R. Konecny, J. Norem, J. Rosenzweig, and J. Simpson, Phys. Rev. Lett. 61, 2756 (1988)

[2] M. C. Thompson, H. Badakov, A. M. Cook, J. B. Rosenzweig, R. Tikhoplav, G. Travish, I. Blumenfeld, M. J. Hogan, R. Ischebeck, N. Kirby, *et. al.*, Phys. Rev. Lett. 100, 214801 (2008)

[3] A. M. Cook, R. Tikhoplav, S. Y. Tochitsky, G. Travish, O. B. Williams, and J. B. Rosenzweig, Phys. Rev. Lett. 103. 095003 (2009)

[4] G. Andonian, O. Williams, X. Wei, P. Niknejadi, E. Hemsing, J. B. Rosenzweig, P. Muggli, M. Babzien, M. Fedurin, K. Kusche, *et. al.*, Appl. Phys. Lett. 98, 202901 (2011)

[5] P. Schoessow, A. Kanareykin, C. Jing, A. Kustov, A. Altmark, J. G. Power, and W. Gai, Proceedings AAC-2008, AIP Conference Proceedings, 1086, 404 (2009)

[6] A. Kanareykin, P. Schoessow, M.Conde, C.Jing, J.G. Power and W.Gai, Proc. Europ. Part. Accel. Conf. EPAC 2006, 2460 (2006)

[7] S. Antipov, C. Jing, A. Kanareykin, P. Schoessow, M. Conde, W. Gai, S. Doran, J. G. Power, Z. Yusof, AIP Conf. Proc, Particle Accelerator Conference, New York, 2074 (2011)

[8] F. Maier, J. Ristein, L. Ley, Phys. Rev. B 64, 165411/1-7 (2001)

[9] P. Muggli, V. Yakimenko, M. Babzien, E. Kallos, and K. P. Kusche, Phys. Rev. Lett. 101, 054801 (2008)

[10] D. C. Nguyen and B. E. Carlsten, Nucl. Instrum. Methods Phys. Res., Sect. A 375, 597 (1996)

[11] M. Ter-Mikaelian, High-Energy Electromagnetic Processes in Condensed Media (Willey-Interscience, New York, 1972)

[12] H. Grote, C. Iselin, The MAD program (methodical accelerator design): version 8.10; user reference manual (Geneva, CERN, 1993)

[13] L. Xiao, W. Gai, and X. Sun, Phys. Rev. E 65, 016505 (2001)

[14] S. Antipov, C. Jing, M. Fedurin, W. Gai, A. Kanareykin, K. Kusche, P. Schoessow, V. Yakimenko, A. Zholents, Accepted to Phys. Rev. Lett. (2012) (arXiv:1111.7291v2)